*Astro2010 Science White Paper (submitted to both CFP and GCT)*
# Spin and Relativistic Phenomena Around Black Holes
L. Brenneman, J. Miller, P. Nandra, M. Volonteri, M. Cappi, G. Matt, S. Kitamoto, F. Paerels, M. Mendez, R. Smith, M. Nowak, M. Garcia, M. Watson, M. Weisskopf, Y. Terashima, Y. Ueda

**I : Probing the formation and growth of black holes using spin**

Ever since the seminal work of Penrose (1969) and Blandford & Znajek (1977), it has been realized that black hole spin may be an important energy source in astrophysics. X-ray observations are uniquely able to answer: **Does black hole spin play a crucial role in powering relativistic jets such as those seen from radio-loud active galactic nuclei (AGN), Galactic microquasars, and Gamma-Ray Bursts?** Indeed, the radio-loud/radio-quiet dichotomy in the AGN population is usually attributed to differences in black hole spin (with correlations between black hole spin and host galaxy morphology being hypothesized in order to explain why radio-loud AGN occur in early-type galaxies; see Figure 1, drawn from Sikora et al. 2007).

The importance of black hole spin goes beyond its role as a possible power source. The spin of a black hole is a fossil record of its formation and subsequent growth history. The details differ somewhat between stellar-mass and supermassive black holes. Given that stellar-mass black holes found in Galactic Black Hole Binaries (GBHBs) are many times more massive than their companion stars, they are unlikely to have changed their spin appreciably since they were formed. To see this, note that a black hole must accrete an appreciable fraction of its original mass in order to significantly change its spin. In the case of GBHBs with low-mass stellar companions, even the accretion of the entire companion star will only change the spin by a small fraction. On the other hand, in GBHBs with high-mass stellar companions, even Eddington-limited accretion will only grow the black hole by a small amount before the high-mass companion explodes. **Thus, the spin of a stellar-mass black hole is a direct relic of the dynamics within the supernova formation event.**

In contrast, **the spin of a supermassive black hole encodes the growth history of the hole and, particularly, the role of mergers versus accretion during the final doubling of the hole's mass** (see Figure 2 drawn from theoretical work by Berti & Volonteri 2008). Successive mergers produce a population of black holes that are spinning at a moderate rate (dimensionless spin parameter a~0.7, where $a = cJ/GM^2$), whereas powerful (quasar-phase) accretion events will produce very rapidly rotating black holes (a>0.9). Yet another possibility is that most black holes have been grown via many short-lived, uncorrelated accretion events, in which case slowly rotating black holes (a<0.5) would be expected (King & Pringle 2006).

Despite its importance, we are only now gaining our first tantalizing glimpses of black hole spin in a few objects. Unlike measurements of black hole mass, spin measurements require us to examine observables that originate within a few gravitational radii of the black hole. A powerful probe of this region that can be applied to black hole systems of all masses is obtained through the study of relativistically-broadened spectral features that are produced in the surface layers of the inner accretion disk in response to irradiation by the hard X-ray source (Tanaka et al. 1995). The strongest feature in this reflection spectrum is the iron-K$\alpha$ line (see Figure 3). For moderate accretion rates (between ~1-30% of the Eddington rate), we expect the iron line emitting part of the disk to extend down to, but be truncated by, the innermost stable circular orbit (ISCO) of the black hole potential (see Reynolds & Fabian 2008 for numerical simulations that support this assertion). This ISCO-truncation imparts a spin-dependence to the reflection spectrum; black holes with higher (prograde) spin have an ISCO at smaller radius and hence the maximum redshift experienced by the reflection spectrum and the iron line is increased.

High signal-to-noise spectra across a wide band-pass are required to obtain robust spin measurements. The effects of spin on the disk reflection spectrum are not subtle, but the disk spectrum must be decomposed from other complexity in the spectrum such as continuum curvature or the effects of photoionized absorbers. For this reason, current studies (with *XMM-Newton* and *Suzaku*) have been limited to a handful of GBHBs (Miller et al. 2009) and one AGN (MCG-6-30-15; Brenneman & Reynolds 2006).

**The International X-ray Observatory (IXO) will make measurements of black hole spin a matter of routine, revolutionizing our knowledge of supermassive black hole spin evolution as well as relativistic jet formation and power.** In addition to the superior throughput in the iron-K band, the high spectral resolution in the 0.5-10 keV band coupled with its hard X-ray sensitivity will enable the disk spectrum to be decomposed from other complexities in the spectrum in a completely unambiguous manner. By targeting known AGN (such as those found in the hard X-ray survey obtained from the *Swift/BAT*), IXO will measure the spin of 200-300 supermassive black holes in the local ($z<0.2$) Universe and a handful (5-10) of supermassive black holes out to $z\sim1$. IXO will also easily determine the spin of every accessible GBHB in the Galaxy or the Magellanic Clouds that enters into outburst during the mission lifetime. Since this study is conducted with electromagnetic tracers of black hole spin (rather than gravitational waves), it will be straightforward to study other aspects of these systems. **In short, we will be able to place the role of black hole spin into context, examining the consequences of spin on radio jet power and host galaxy morphology.**

The disk reflection methodology outlined here is powerful since it can be applied uniformly to black holes across the whole range of masses. However, there are other mass-specific techniques that lead to spin constraints. This can be used to provide crucial consistency checks. In GBHBs, detailed examination of the thermal X-ray emission from the accretion disk gives another measurement of the ISCO and hence the spin (Shafee et al. 2006). GBHBs also occasionally display high-frequency quasi-periodic oscillations (HFQPOs) which provide a third (albeit model-dependent) constraint. Finally, a fourth method is provided by polarimetry; the polarization angle of disk emission rotates with energy due to GR effects (Dovciak et al. 2008).

Completely novel consistency checks are possible for the brightest AGN. Rapid iron line variability is expected from orbiting structures within the disk and/or the reverberation of X-ray flares across the disk (see Section 2 below). Both of these phenomena have well defined spin-dependence. Furthermore, the first AGN-QPO has recently been reported (Gierliński et al. 2008) and opens up the possibility of timing-based measurements of supermassive black hole spin.

The IXO census of spins will revolutionize black hole astrophysics. The spins of stellar-mass black holes in GBHBs are natal and hence give a direct window into the birth of these objects. This, in turn, provides a glimpse into the workings of the most powerful explosions in the Universe – at least some stellar mass black holes are believed to be born in long Gamma-Ray Bursts. For AGN, the IXO spin census will allow us to determine the distribution of black hole spins as functions of host galaxy type and redshift. Comparisons with detailed theoretical calculations will determine whether supermassive black hole growth has been dominated by accretion or by mergers. On all mass scales, correlations between black hole spin and the presence of relativistic jets will provide a clean test of the hypothesis that a rapidly rotating black hole is the basic power source for these jets.

**II : Strong gravitational physics close to black holes**

Black holes provide the ultimate laboratory for studying strong-field gravitational physics. Of course, the most compelling theory of gravity to date is General Relativity (GR). For weak gravitational fields, GR has passed precision tests (within the framework of Parameterized Post Newtonian theory in which all variants of GR are encapsulated), but GR remains essentially untested in the strong-gravity regime. **Given that GR is one of the foundations of modern physics, it is important to explore its predictions in many independent and unbiased ways.**

In luminous black hole systems, the accretion flow is in the form of a thin disk of gas orbiting the black hole. To a very good approximation, each parcel of gas within the disk follows a circular test-particle orbit (e.g. Armitage & Reynolds 2003). **This geometrical and dynamical simplicity makes accretion disks useful for probing the black hole potential and, hence, the predictions of GR.**

Current studies by *XMM-Newton* and *Suzaku* clearly show the broadening and skewing of the disk reflection features in both AGN and GBHBs predicted by GR (due to a combination of the relativistic Doppler shift and gravitational redshift; for a review, see Miller 2007). In a small number of AGN, observations already hint at the power of orbit-by-orbit traces using emission lines (Figure 4). However, most studies must integrate for many orbits of the accretion disk in order to collect enough photons to define the disk reflection spectrum. This time-averaging removes much of the dynamical information and, in particular, makes it impossible to simultaneously measure black hole spin and search for deviations from the predictions of GR.

With its superior throughput, IXO will sweep away this degeneracy. IXO will enable detection of iron line variability on sub-orbital timescales in approximately 20-30 AGN. Any non-axisymmetry in the emission of the iron line (e.g. associated with the expected turbulence in the disk) will lead to a characteristic variability of the iron line, with "arcs" being traced out on the time-energy plane (Figure 5). GR makes specific predictions for the form of these arcs, and the ensemble of arcs can be fitted for the mass and spin of the black hole, as well as the inclination at which the accretion disk is being viewed. Deviations from the predictions of GR can be sought by searching for *apparent* changes of the inferred mass and/or spin with radius in the disk.

A second kind of emission line variability will occur due to the reverberation (or "light echo") of X-ray flares across the accretion disk. Observing this behavior is critical, as it enables an absolute determination of the size scale (GR shifts giving this only in terms of the gravitational radius). Reverberation observations offer unambiguous proof of the origin of the X-ray lines as reflection features, allowing us to map the geometry of the X-ray source and inner accretion flow. The path and travel time of photons close to the black hole is also strongly affected by space-time curvature and frame-dragging. In systems with very rapidly rotating black holes, the region of the accretion disk capable of producing line emission can extend down almost to the event horizon, so we can probe time-delays along photon paths that pass close to the horizon. These photon paths create a low-energy, time-delayed "tail" in the GR reverberation transfer function. The nature of this tail is insensitive to the location of the X-ray source but is highly sensitive to the spacetime metric, so characterizing this will provide another potential test of GR, this time based on photon orbits rather than matter orbits.

The reverberation of individual flares will be accessible to IXO in the brightest few AGN. However, reverberation will be statistically detected in many more AGN and GBHBs via the use of Fourier techniques aimed at detecting the lag between the driving continuum emission and the strongest fluorescent emission lines.

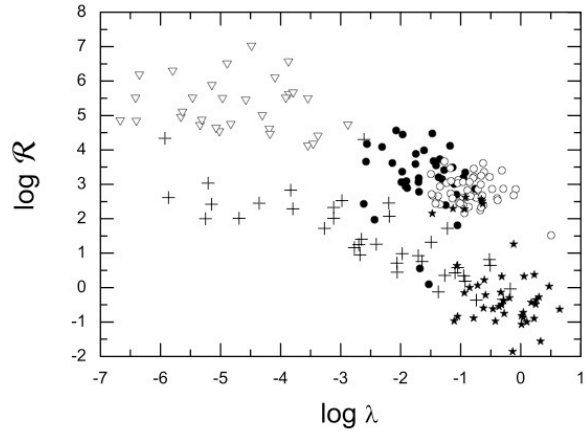

Figure 1 : **The spin-paradigm in AGN** : Radio/optical (5GHz/440nm) flux ratio $\mathcal{R}$ against Eddington ratio $\lambda$ for a sample of AGN (broad-line radio galaxies are filled circles, radio-loud quasars are open circles, Seyfert galaxies and LINERs are crosses, FRI radio galaxies are open triangles, PG quasars are filled stars). The radio-loud/radio-quiet dichotomy is revealed via the existence of two distinct tracks on this plot. The spin-paradigm holds that the upper branch (exclusively early type galaxies) corresponds to rapidly spinning SMBHs whereas the lower branch (all galaxy types) corresponds to slowly spinning SMBHs. **IXO observations of these sources can directly test this paradigm.** (Figure taken from Sikora, Stawarz & Lasota 2007).

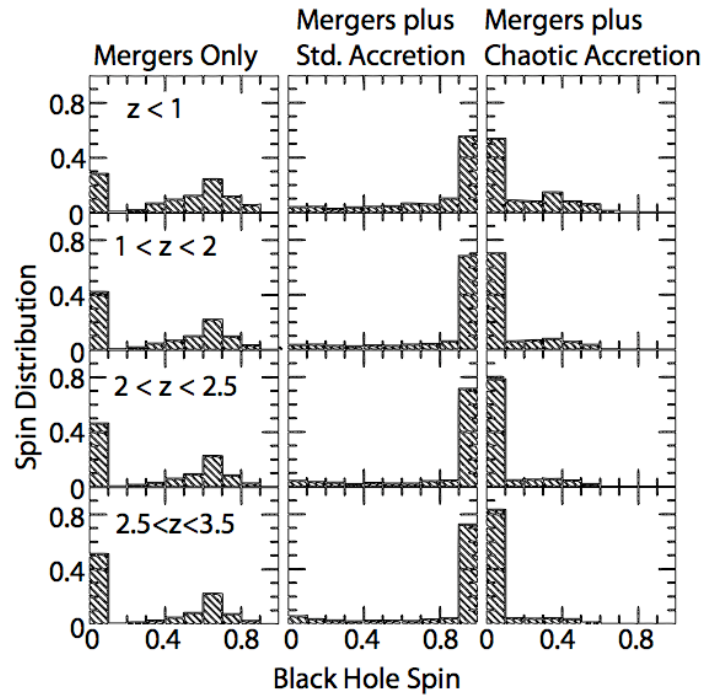

Figure 2 : **Spin as a probe of SMBH growth history**. Left panels show a predicted distribution of SMBH spin in a scenario where seed black holes are formed at high redshift after which the SMBH population evolves purely via SMBH-SMBH mergers. The middle panels show a predicted spin distribution resulting from mergers plus standard (disk) accretion. The right panels show the distribution when appreciable SMBH disk-accretion accompanies merger events. **The spin-distribution is a powerful discriminant between growth histories that may form identical mass-functions**. (Figure adapted from Berti & Volonteri 2008).

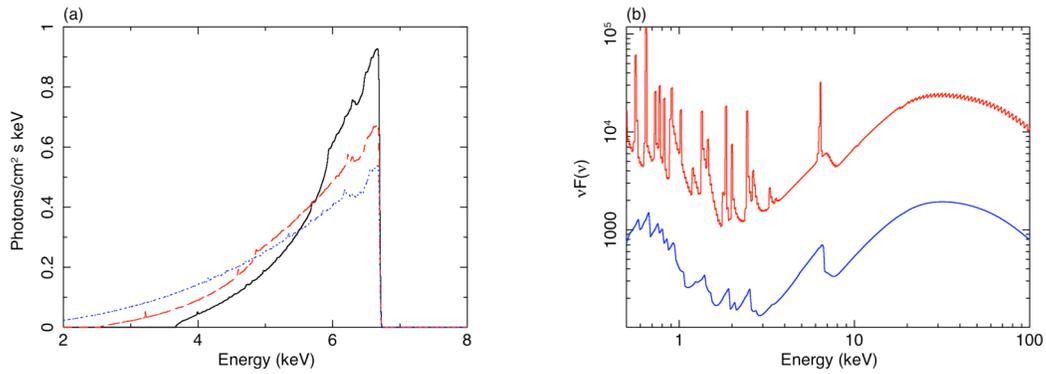

Figure 3: **Relativistically-broadened disk reflection features**. (a) ISCO-truncated iron line profiles for spins of a=0 (black), a=0.7 (red) and a=0.998 (blue). (b) An ionized reflection spectrum from before (top) and after (bottom) the relativistic smearing effects (for a=0.7) are incorporated. **The iron line is a powerful probe of black hole spin**.

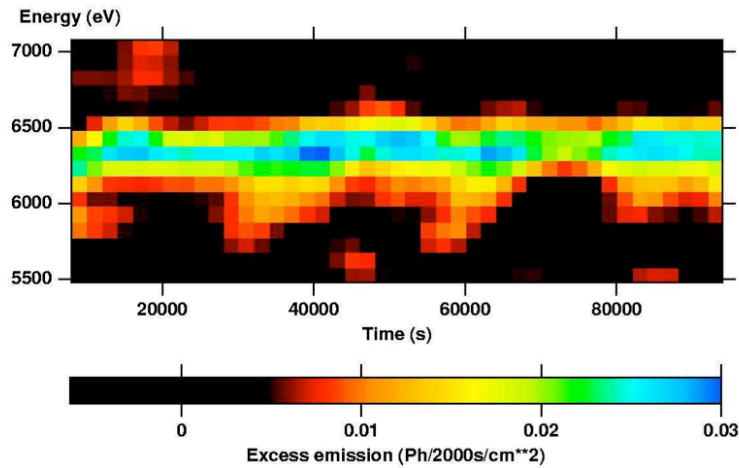

Figure 4: Current observations of AGN are just able to search for emission line variations on the orbital timescale at the ISCO. The plot above depicts Doppler shifts in an iron emission line consistent with Keplerian orbits at the ISCO in the Seyfert AGN NGC 3516. The saw-tooth pattern is exactly that expected for orbital motion in the innermost relativistic regime around black holes. The sensitivity of IXO will make it possible to explore these timescales in fine detail, revealing the nature of the strong-field gravitational potential. (Figure taken from Iwasawa, Miniutti, & Fabian 2004.)

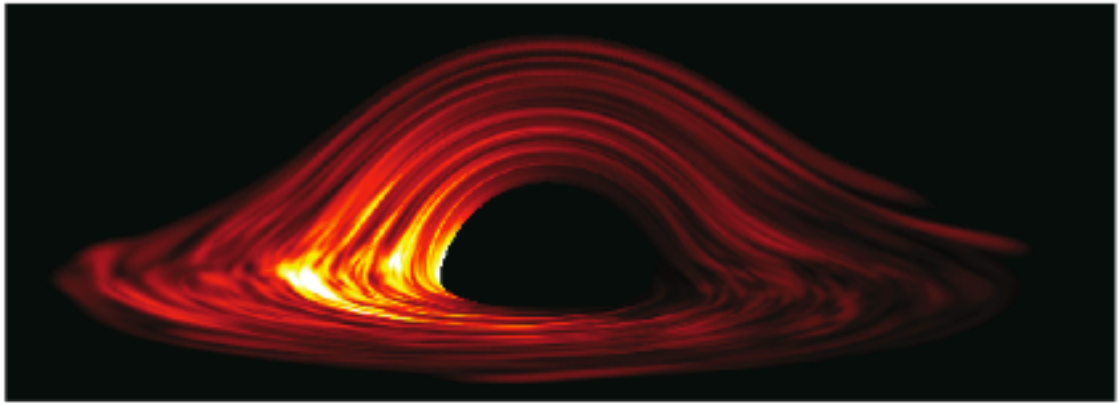

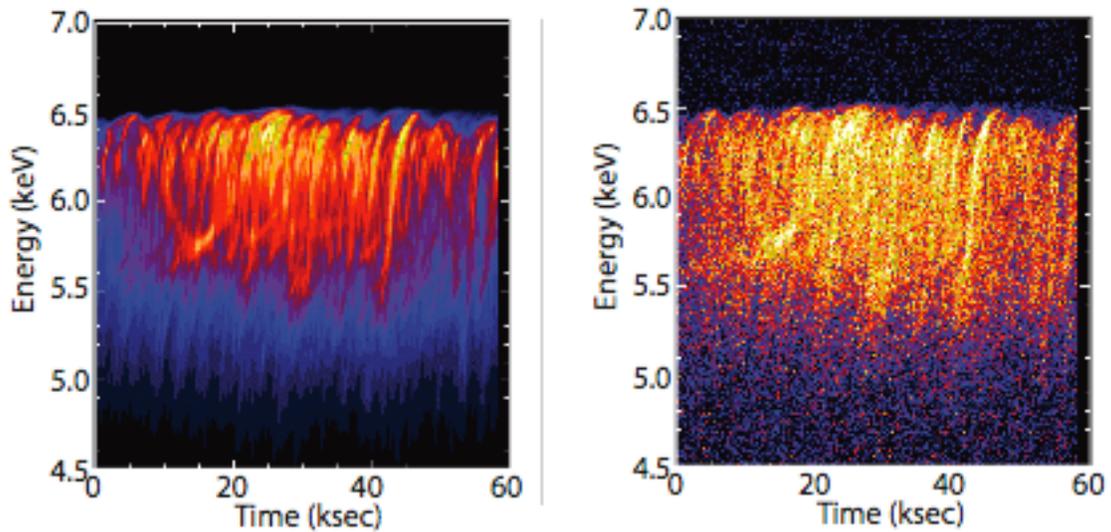

Figure 5: **How does matter behave close to a black hole?** Top panel: Simulated image of a highly-inclined MHD turbulent accretion disk around a Schwarzschild black hole. Bottom left panel : The iron line in the time-energy plane from a turbulent disk viewed at an inclination of 20 degrees. **Note the arcs in the trailed-profile that can be directly mapped to test-particle like orbital motion of hot-spots in the disk.** Bottom right panel : Result of a simulated IXO observation, assuming a $3 \times 10^7$ $M_{sun}$ black hole and a 2-10keV flux characteristic of a bright AGN. (Figure adapted from Armitage & Reynolds 2004).
A video of an MHD simulation of a turbulent disk at 30 degrees inclination, and the time-resolved iron lines that it would produce, can be seen at:
http://ixo.gsfc.nasa.gov/documents/resources/posters/aas2009/brenneman_plunge_30i.avi